\documentclass{iaus}
\usepackage{graphicx}

\newcommand{\pun}[1]{\mbox{\rm\,#1}} 
\newcommand{\Teff}{\ensuremath{T_{\mathrm{eff}}}}
\newcommand{\logg}{\ensuremath{\log g}}

\title[Prospects for simulations of brown dwarf photospheres]
{Prospects of using simulations to study the photospheres of brown dwarfs}

\author[H.-G. Ludwig]   
{Hans-G{\"u}nter Ludwig$^1$}

\affiliation{$^1$CIFIST, GEPI, Observatoire de Paris-Meudon, 
92195 Meudon, France\break email: Hans.Ludwig@obspm.fr}

\pubyear{2007}
\volume{239}  
\pagerange{xxx--xxx}
\date{?? and in revised form ??}
\jname{Convection in Astrophysics}
\editors{F. Kupka, I.W. Roxburgh \& K.L. Chan, eds.}
\begin{document}

\maketitle

\begin{abstract}
We discuss prospects of using multi-dimensional time-dependent
  simulations to study the atmospheres of brown dwarfs and extrasolar
  giant planets, including the processes of convection, radiation, dust
  formation, and rotation. We argue that reasonably realistic simulations are feasible,
  however, separated into two classes of local and global models. Numerical
  challenges are related to potentially large dynamic ranges, and the treatment
  of scattering of radiation in multi-D geometries.
\end{abstract}


\section{Introduction}
The increasing number of brown dwarfs and extrasolar planets of spectral class
L and later discovered by infrared surveys and radial velocity searches has
spawned a great deal of interest in the atmospheric physics of these objects.
Their atmospheres are substantially cooler than, e.g., the solar atmosphere,
allowing the formation of molecules, or even liquid and solid condensates --
in astronomical parlance usually referred to as ``dust''. Convection is a
ubiquitous phenomenon in these atmospheres shaping their thermal structure and
the distribution of chemical species.  Hydrodynamical simulations of solar and
stellar granulation have become an increasingly powerful and handy instrument
for studying the interplay between gas flows and radiation. In this paper we
discuss the prospects of developing similar \textit{multi-dimensional} and
\textit{time-dependent} simulations of very cool atmospheres. The most
important additional process -- in view of previous developments for hotter
atmospheres -- that one needs to tackle is the dust formation coupled to the
hydrodynamic transport processes and radiative transfer.

In the following, we shall take a slightly broader point of view than just
considering brown dwarf (hereafter BD) atmospheres and include also the atmospheres of
extrasolar giant planets (hereafter EGPs) in the discussion since their atmospheric
dynamics is controlled by similar processes as brown dwarfs atmospheres.

As we shall later see, computational limitations do not allow to address the
problem of the atmospheric dynamics as a whole, i.e., studying the global
atmospheric circulation together with convective flows taking place at
relatively small spatial scales given by the the pressure scale height at the
stellar surface. Consequently modelers have addressed the problem of the
local and global circulation separately. Since many BDs and EGPs are
rapid rotators rotation constitutes another important process which one needs
to take into consideration for the global circulation.

We would like to emphasize that we are discussing the perspectives for multi-D
time-dependent simulations. Considerable insight has already been gained by
the development of one-dimensional, hydrostatic, time independent model
atmospheres (hereafter ``standard'' models) for L- and T-type objects (e.g.,
\cite{Burrows06,Tsuji02,Allard01,Ackerman01}), and we draw from this work.
Simulations will augment our understanding of BD and EGP atmospheres by adding
information about the detailed cloud meteorology on the local scale of
convective cells, as well as on the scale of the global wind circulation
pattern. This includes further characterization of the effects of irradiation
in close-in EGPs.  The simulation of the atmospheric dynamics might also add
to our knowledge about local dynamo action in substellar objects, and acoustic
activity contributing to the heating of chromospheres.

\section{Micro-physical input}

In order to perform realistic simulations micro-physical input data must be
available -- radiative opacities, equation-of-state (EOS), and a kinetic model
describing the formation of dust grains. The requirements are similar to those
for standard models, and consequently in simulation work one can usually take
recourse to the descriptions developed for 1D models -- largely on the same
level of sophistication. In all three before-mentioned areas substantial
progress has been made over the last decade, spawned by the discovery of the
first brown dwarf and EGP in 1995. In particular, since the early work of
\cite{Rossow78} kinetic models describing the nucleation, growth, and
evaporation of dust grains under conditions characteristic of brown dwarf
atmospheres have been developed, see \cite{Helling04} and references therein.
Hence, the present input data allow to set-up simulations on a sufficiently
realistic level.

\section{Time scales: convection, radiation, dust, rotation, \&\ numerics}

\begin{figure}[t]
\begin{center}
\includegraphics[width=0.7\textwidth]{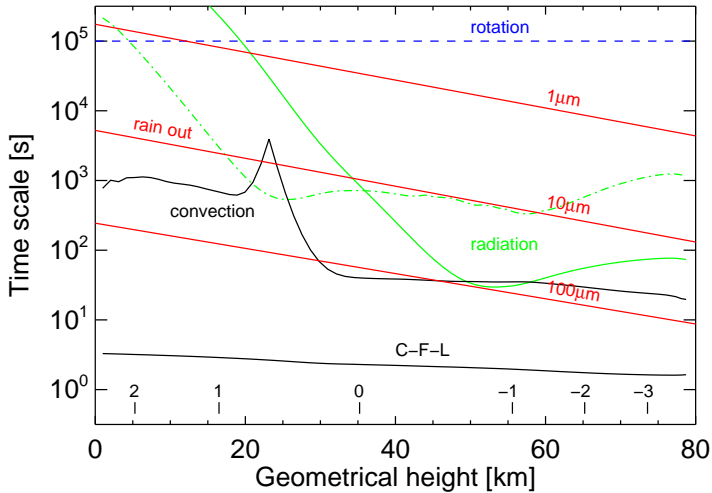}
\caption{Characteristic time scales of various processes in a brown dwarf
  atmosphere of \Teff=1800\pun{K} as a function of geometrical height. The
  tick marks close to the abscissa indicate the ($\log$ Rosseland) optical
  depth. For details see text.}\label{f:ctimes}
\end{center}
\end{figure}

To obtain insight into potential challenges one faces in simulations of the
dynamics of brown dwarf and EGP atmospheres it is illuminating to take a look
at the characteristic time scales of the governing physical processes.
Figure~\ref{f:ctimes} depicts these time scales in a representative brown
dwarf model atmosphere at \Teff=1800\pun{K} and \logg=5.0 of solar chemical
composition. The model comprises the stellar photosphere and the uppermost
layers of the convective stellar envelope. Since it is expected that the cloud
decks are located in vicinity of the boundary of the convective envelope (in
this model located at a geometrical height of 23\pun{km}) it also contains the
layers in which the dust harboring layers are expected. We emphasize that the
model structure is taken from an experimental hydrodynamical simulation in
which dust formation was not taken into account. Since here we are interested
in order of magnitude estimates only this is not a critical issue.

In figure~\ref{f:ctimes} the line labeled ``C-F-L'' computed as the sound
crossing time over a pressure scale height depicts the upper limit of the time
step which is allowed in an explicit hydrodynamical scheme due to the
Courant-Friedrichs-Levy stability criterion. Depending on the actual
resolution of the numerical grid this number may be one to two orders of
magnitude smaller than indicated. The line labeled ``convection'' depicts the modulus
of the Brunt-Vai{\"a}s{\"a}l{\"a} period providing a measure of the time scale
on which the convective flow evolves. Two lines labeled ``radiation'' indicate
the time scale on which radiation changes the thermal energy content of the
gas. The dashed-dotted line is computed using Rosseland mean opacities which
give the correct behavior in the optical thick layers, the solid line is based
on Planck mean opacities which give a better representation in the optically
thin layers. For the rotational period depicted by the dashed line labeled
``rotation'' we took a representative value close to one day. For three
different dust grain diameters of 1, 10, and 100\pun{$\mu$m} we plotted the
sedimentation (``rain out'') time scale taking as drifting time over a
pressure scale height. The drift velocities were taken from the work of
\cite{Woitke03}. We did not depict the formation time scale of the dust grains
in the figure: for grains of 100\pun{$\mu$m} diameter it is of the same order
as the sedimentation time scale. Consequently, it is unlikely that larger
grains can stay in brown dwarf atmospheres. The formation time scale
becomes rapidly shorter for smaller grains so that they can be considered
being essentially formed in quasi-static phase-equilibrium (see also
\cite{Helling05}.

Computing resources available today typically allow to simulate a dynamical
range in time of $10^4 $\ldots$ 10^5$, and $10^2 $\ldots$ 10^3$ per dimension
(for 3D models) in space. From figure~\ref{f:ctimes} we conclude that it
should be feasible to include convection, radiative transfer effects, and dust
formation in a simulation of a BD/EGP atmosphere. The simultaneous inclusion
of rotation is beyond reach, in particular if one takes into consideration
that one would like to simulate many rotational periods to obtain a
statistically relaxed state. However, the substantial difference between the
time scales on which rotation and convection operate moreover indicates that
rotation is dynamically not relevant for the surface granulation pattern in
BD/EGPs.

A rather strong modeling limitation comes about by the large spatial scale separation
between the typical size of a convective cell and the global scale of a BD or
EGP of about $10^4$, at best reduced to $10^3$ for the case of young, low mass
EGPs. Hence, typical BD/EGP conditions are hardly within reach with 3D models,
and the steep increase of the computational cost with spatial resolution (for
explicite numerical schemes with $(\Delta x)^4$) makes it likely that this
situation prevails during the nearer future. We expect that 3D simulations will
for some time be either tailored to simulate the global meteorology, or will be
restricted to local models simulating the convective flow in detail.

\begin{figure}[t]
\begin{center}
\includegraphics[width=0.7\textwidth]{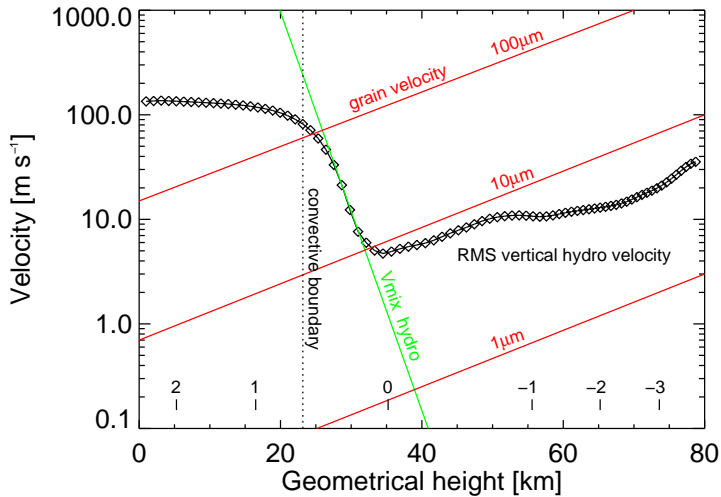}
\caption{Characteristic velocities in brown dwarf atmosphere analogous
  to figure~\ref{f:ctimes}.}\label{f:bdmix}
\end{center}
\end{figure}

Figure~\ref{f:bdmix} illustrates the kinematics of the flow in a local BD
simulation analogous to figure~\ref{f:ctimes}. The horizontal root-mean-square
of the vertical velocity component is depicted by the diamond symbols. The
key-point to note is that the convective motions proper are largely confined to
the convectively unstable layers. The velocities in the convectively stable
layers with $\log\tau<0$ are almost exclusively related to sound waves. As
essentially oscillatory motions they are ineffective for mixing so that they
provide little updraft to keep dust grains aloft in the atmosphere. The green
line is illustrating an estimate of the effective mixing velocity provided by
the convective motions. The decline of the amplitude is rather steep -- in
the test model with a scale height of about $1/3$ of the local pressure scale
height.

Comparing the mixing velocities with typical grain sedimentation velocities
indicates that the kinematics could support cloud decks in the convective zone
and a thin adjacent overshooting layer at its top boundary. Fitting observed
spectra \cite{Burrows06} find a preference for a rather large grain size
of $\approx 100$\pun{$\mu$m} in BD atmospheres. This would make the grain
sedimentation velocities comparable to convective velocities which is
numerically uncritical. More demanding would be small grain sizes. The
distribution of small grains would hinge on the capability of a numerical code
to deal with large velocity ranges. Any non-physical diffusivity in a code can
artificially extend the region over which clouds of small grains could
exist. 

\section{The multi-D story so far}

A number of simulations of BD/EGP atmospheres have been already conducted in
2D and 3D geometry. Here, the problem of the circulation between the day- and
night-side of close-in EGPs (``hot Jupiters'') achieved particular attention.
However, to our knowledge none of the studies has addressed the coupled
problem of hydrodynamics, dust formation, radiation, and rotation, but rather
have focused on different parts of the overall problem. We would like to refer
the interested reader to \cite{Showman02}, \cite{Cho03}, \cite{Burkert05},
including the follow-up works of these groups.

\section{Serendipity}

\begin{figure}[t]
\begin{center}
\includegraphics[width=0.33\hsize]{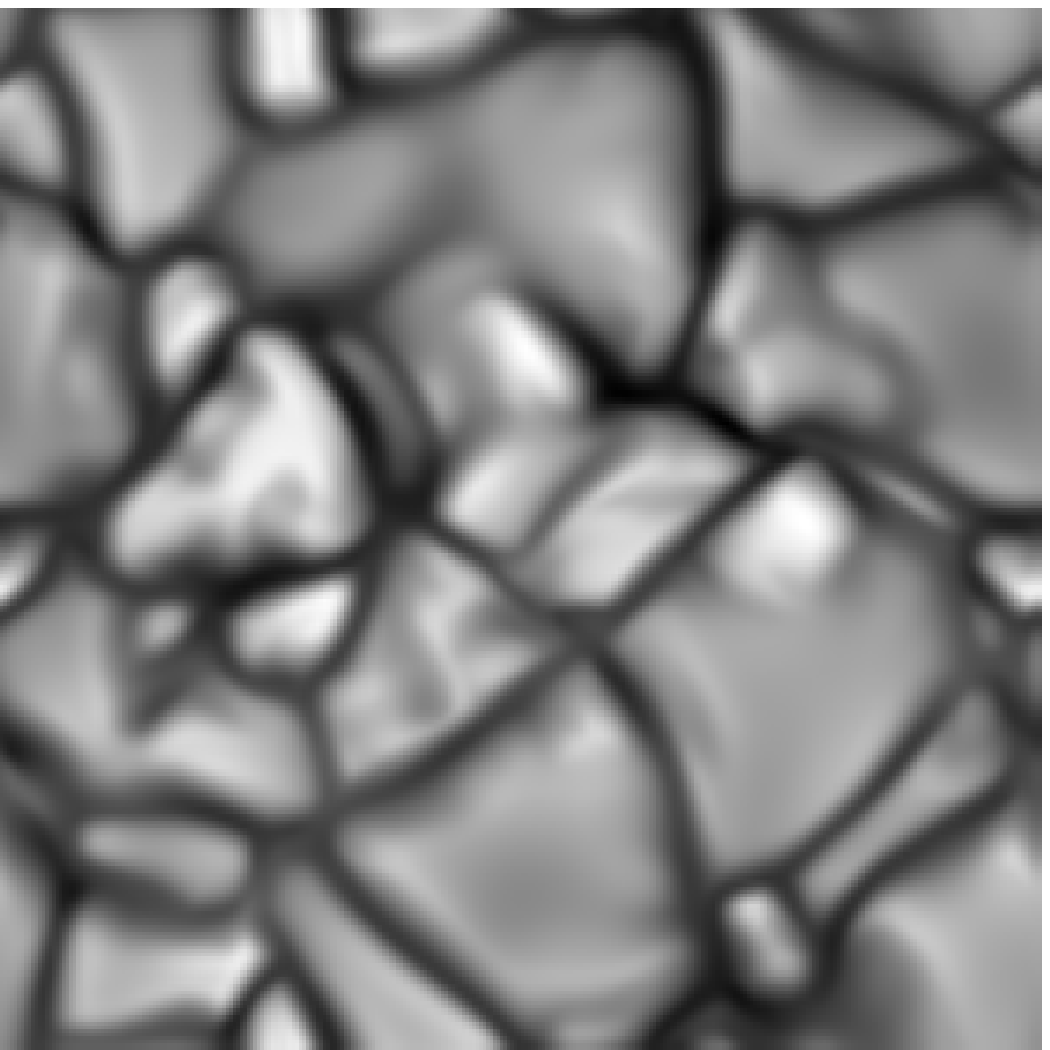}\hfill
\includegraphics[width=0.33\hsize]{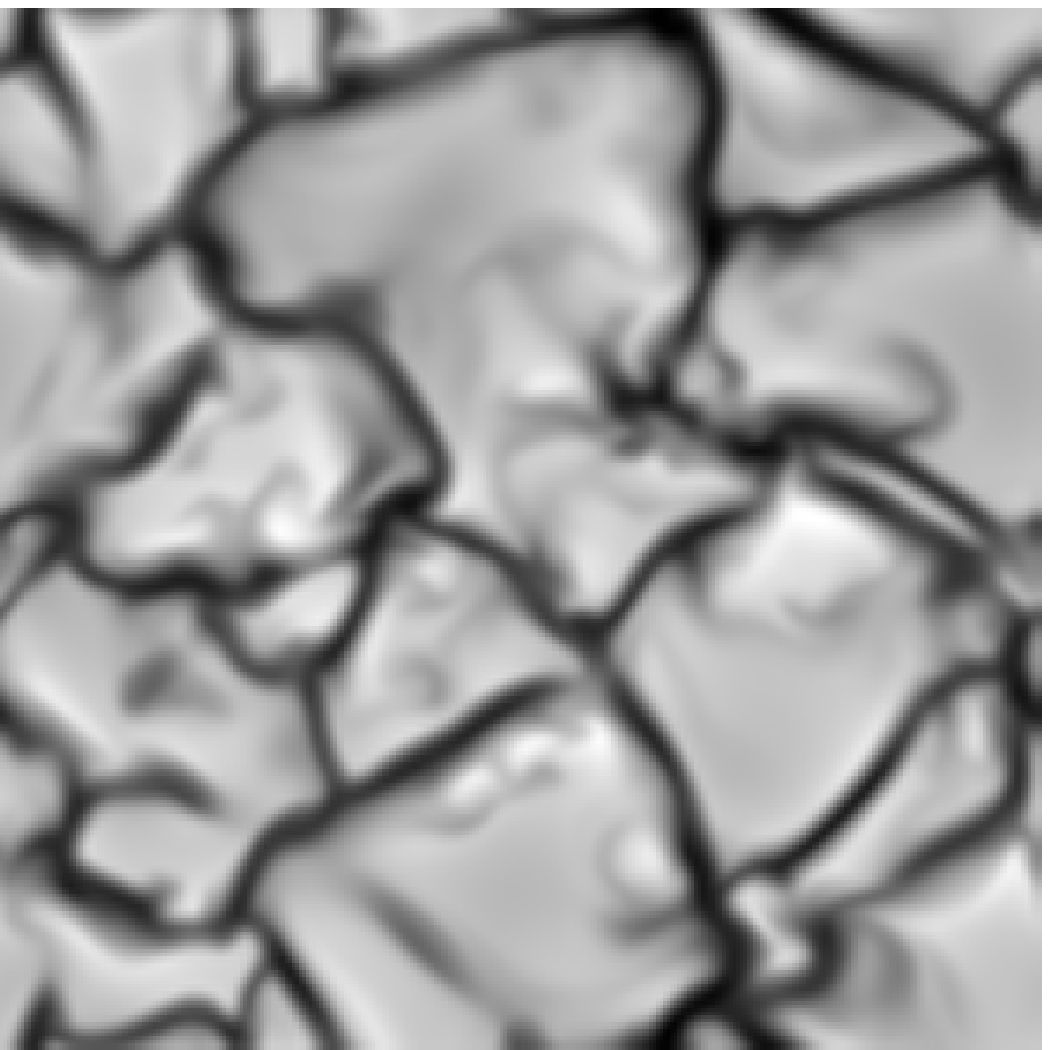}\hfill
\includegraphics[width=0.33\hsize]{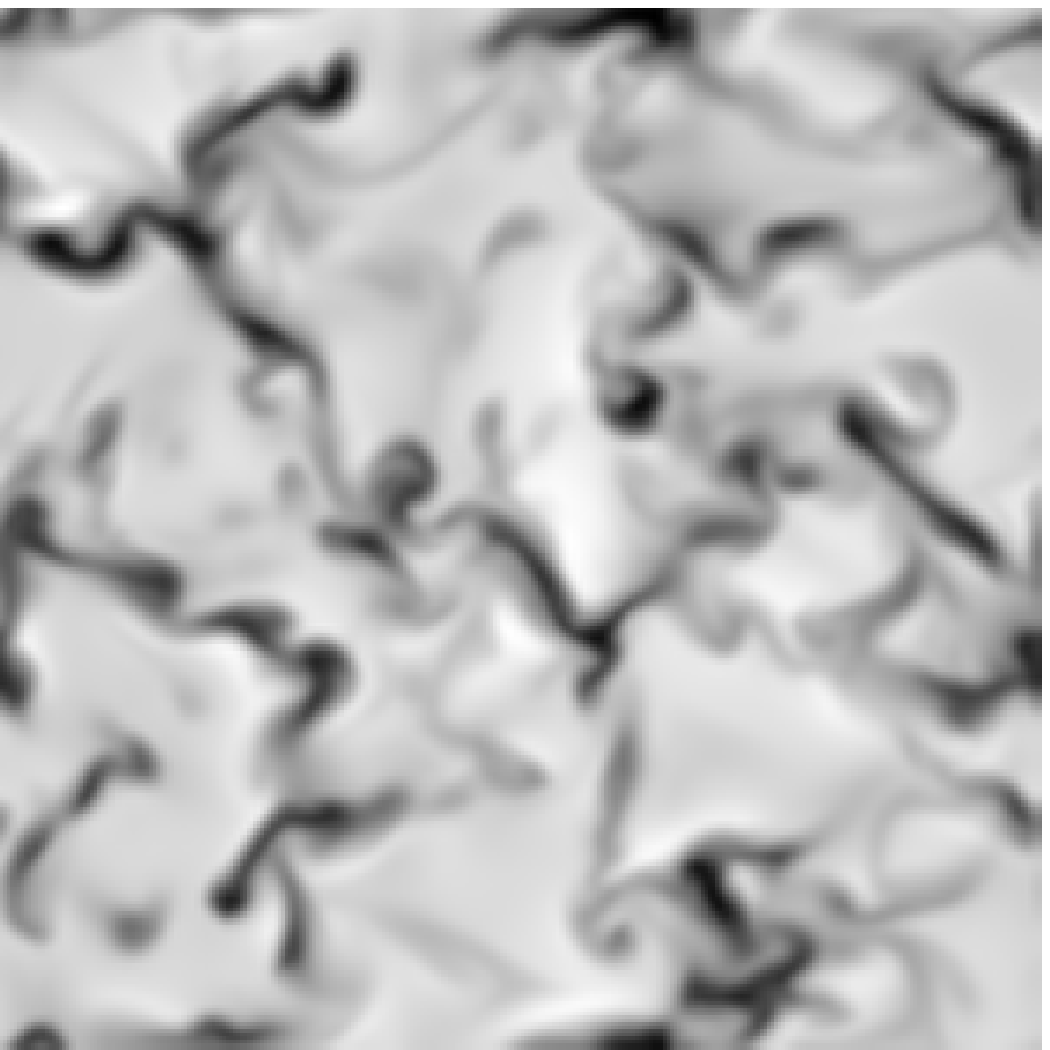}\\[0.1\baselineskip]
\includegraphics[width=0.33\hsize]{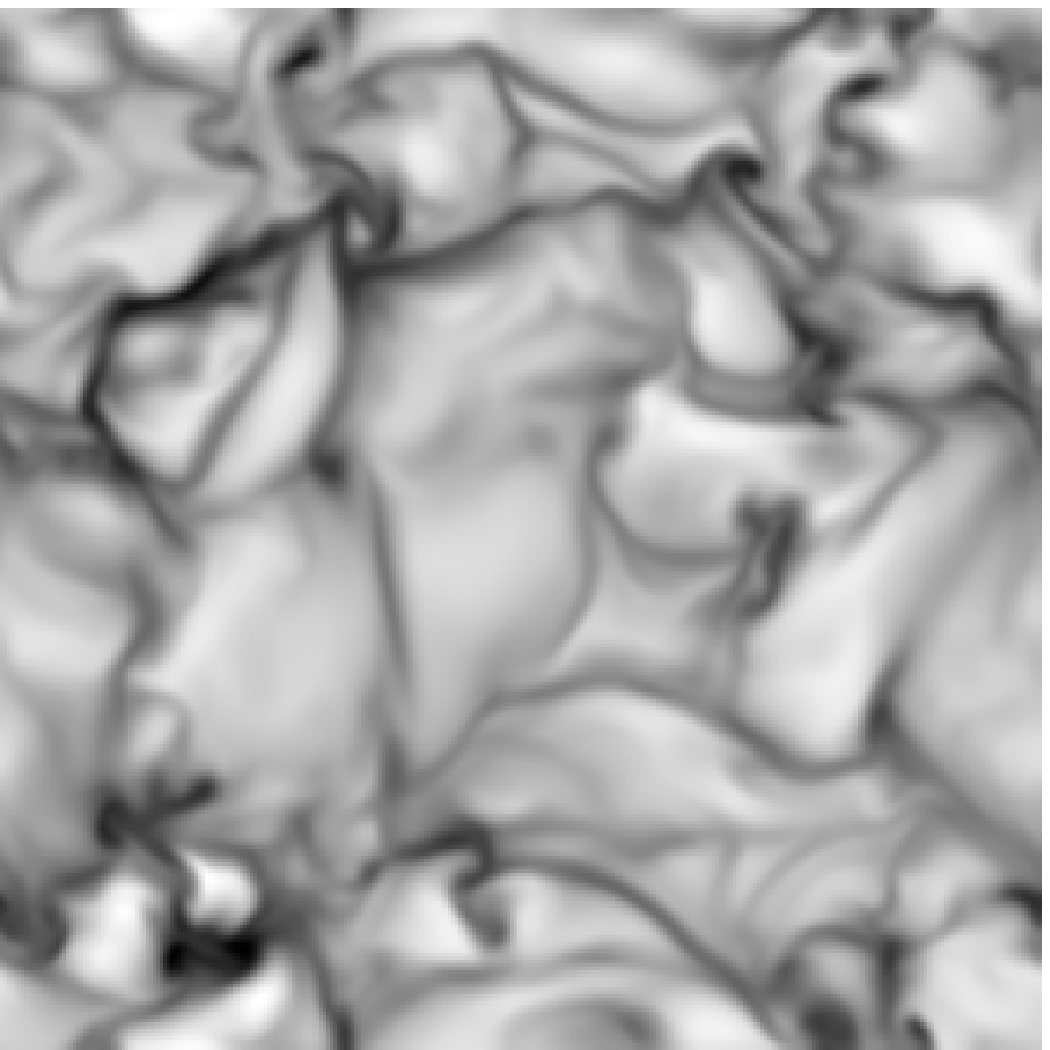}\hfill
\includegraphics[width=0.33\hsize]{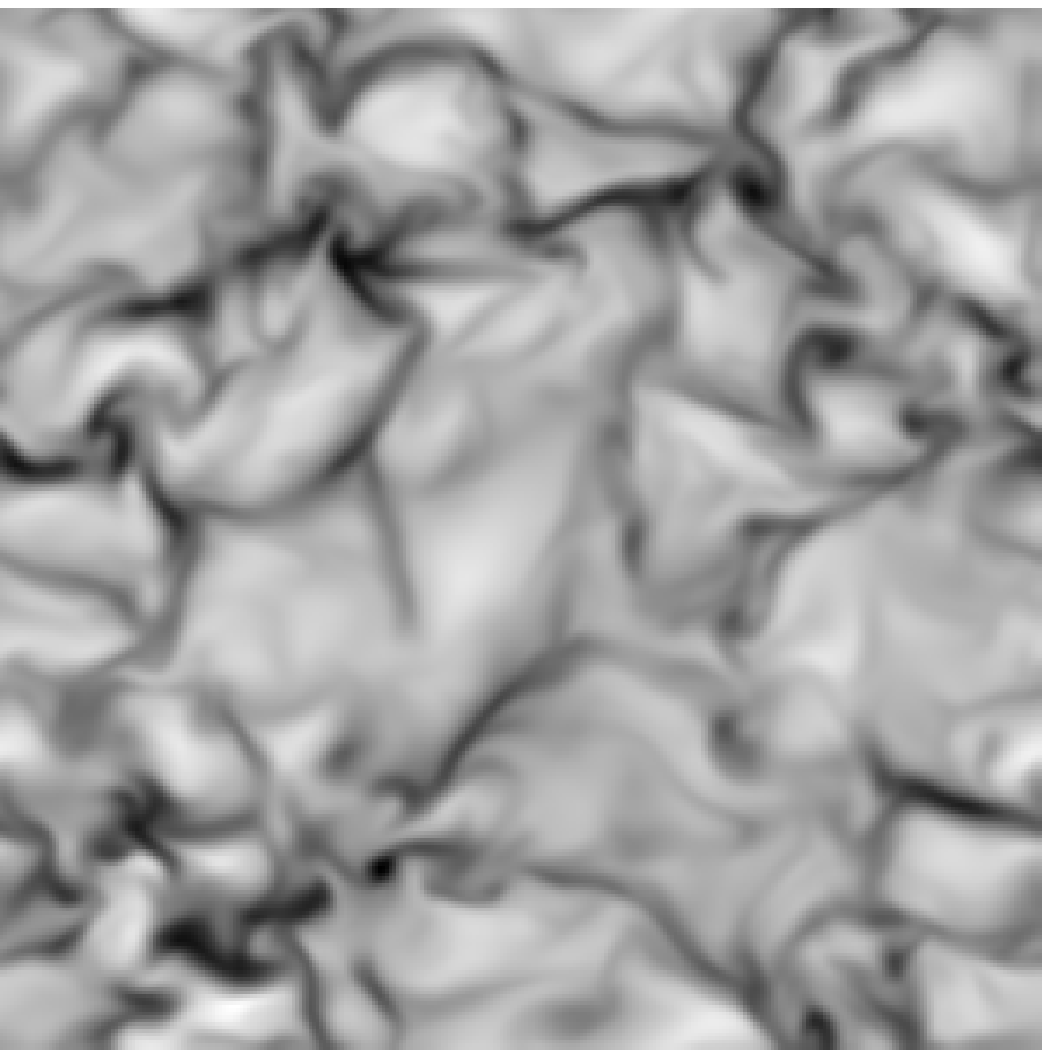}\hfill
\includegraphics[width=0.33\hsize]{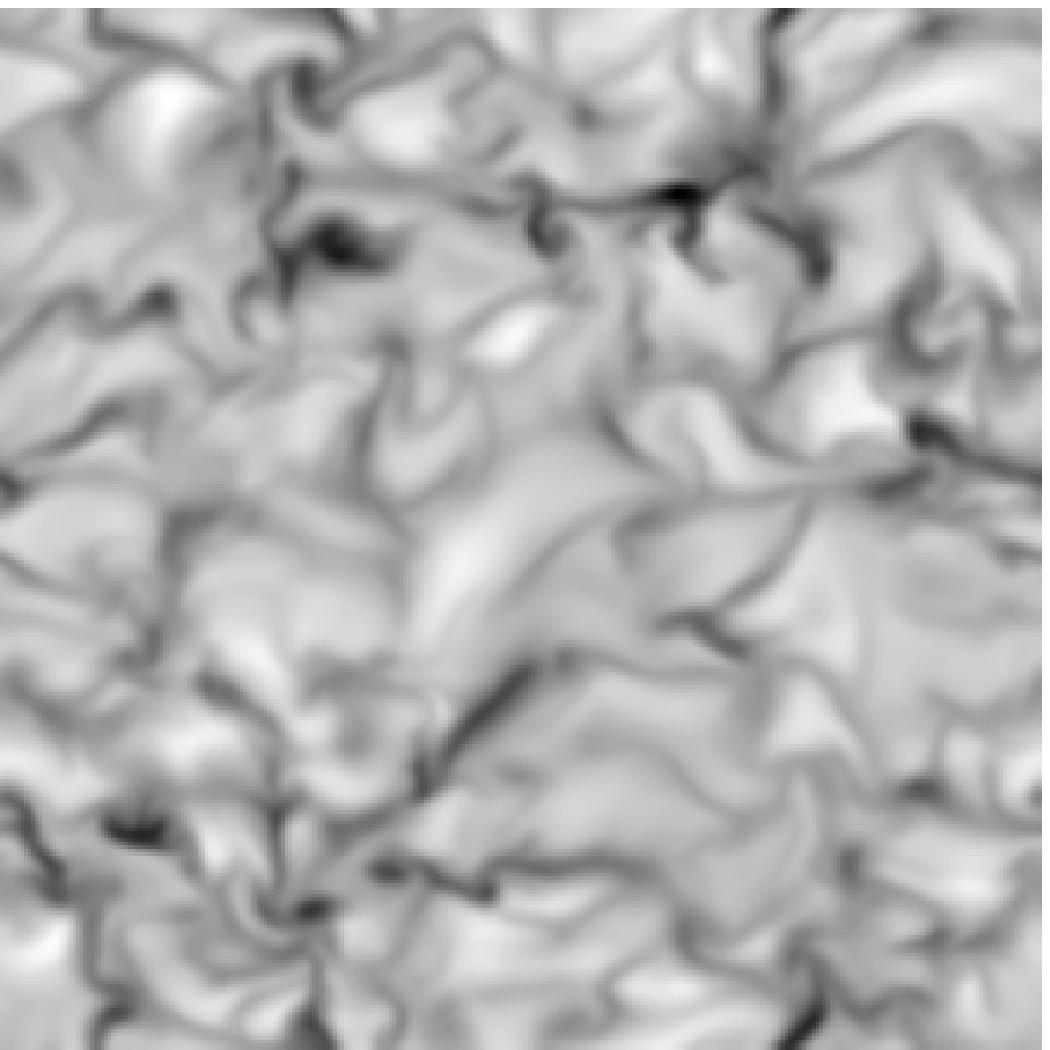}\\
\caption{Grey-scale images of the vertical velocity component of a solar
  hydrodynamical model (top row) and a M-dwarf model (bottom row). From left
  to right, the velocities are depicted at (Rosseland) optical depth unity as
  well as one and two pressure scale heights below that level in the
  respective models. The absolute image scales are $5.6\times 5.6$\pun{Mm$^2$}
  for the solar and $0.25\times 0.25$\pun{Mm$^2$} for the M-dwarf
  model.}
\label{f:flowfields}
\end{center}
\end{figure}

In the previous sections we summarized expectations about the insights one
might gain, and challenges one might face when trying to construct multi-D models
for BD/EGP atmospheres. We added figure~\ref{f:flowfields} as a reminder that
of course the unforeseen results are the most interesting ones.
Figure~\ref{f:flowfields} illustrates a slight but distinct change of the
granulation pattern between the familiar solar granulation and granulation in
an M-dwarf. E.g., similarly and perhaps more drastically the formation of dust
in BD/EGP atmospheres might modulate the convective dynamics in unexpected
ways -- who knows.

\vspace{-\baselineskip}

\section{Conclusions}

Reasonably realistic local or global models of brown dwarf and extrasolar
giant planet atmospheres coupling hydrodynamics, radiation, dust formation,
and rotation are numerically in reach at present. However, ``unified'' models
spanning all spatial scales from the global scale down to scales resolving the flow
in individual convective cells are stretching the computational demands beyond
normally available capacities. Hence, we expect that a separation between
local and global models will prevail during the nearer future. Whether this will
turn out to be a severe limitation remains to be seen. While apparently subtle
we would like to point to the solar dynamo problem where the still not fully
satisfactory state of affairs might be related to the lack of the inclusion of
small enough scales when modeling the global dynamo action. In BD/EGP
atmospheres it is perceivable that the local transport of momentum by
convective and acoustic motions might alter the global flow dynamics -- in the
simplest case by adding turbulent viscosity.

If the sizes of dust grains in BD/EGP atmospheres turn out to be small, and
the grains consequently exhibit low sedimentation speeds, numerical simulations
must have the ability to accurately represent the large dynamic range between
grain and convective/acoustic velocities.  Overly large numerical diffusivities
artificially enlarge the height range over which cloud decks can persist.

Standard model atmospheres are treating the wavelength-dependence of the
radiation field commonly in great detail which is not possible in the more
demanding multi-D geometry of simulation models. An approximate multi-group
treatment of the radiative transfer has been developed for simulations of
stellar atmospheres which has also been proven to provide reasonable accuracy
at acceptable computational cost in cooler (M-type) atmospheres. We expect that
the scheme also works for even cooler atmospheres. However, one simplification
usually made is treating scattering as true absorption.  Depending on the
specific dust grain properties this approximation might need to be replaced by
a more accurate treatment of scattering. Hence, another challenge a modeler
might face is to device a computationally economic scheme to treat scattering
in the time-dependent multi-D case.

\vspace{-\baselineskip}

\begin{discussion}

\discuss{C. Helling}{%
Your wish list implies that no progress has been made in the brown dwarf
modeling. Additionally, I am convinced that we will need to work on both
sides: on 1D models which are fast and applicable, not only on 3D models
though they will play an important role.}

\discuss{Ludwig}{My wish list was intended as overall collection of things we
  would like to understand about brown dwarf atmospheres. Progress related to
  the various points has indeed already been made. Concerning the mutual role
  of 1D and 3D models, I fully agree. 3D models should address crucial
  aspects that are in principle not accessible in 1D. Insight
  emerging from 3D models should then be transferred to 1D models.}

\discuss{F. Kupka}{%
Considering the complexity of molecular opacity I am actually
surprised how robust the opacity binning seems to be.}

\discuss{Ludwig}{%
Tests have been performed for M-type stars where the effect of many millions
  of -- primarily molecular -- lines is captured quite well. As for brown
  dwarfs: the dust opacity has a rather smooth functional dependence on
  wavelength. Hence, it should be easy, but scattering is a problem.}

\discuss{I.W. Roxburgh}{%
You said overshooting was small, could you quantify this in
terms of local scale height?}

\discuss{Ludwig}{The velocity amplitude declines exponentially with a scale
  height of about 1/3 of the local pressure scale height. For comparison: in
  solar models the scale height of decline is about six times larger. However,
  keep in mind that the hydrodynamical model presented here is experimental,
  in particular does not include any effects of dust formation.}
\end{discussion}

\end{document}